\documentclass[prl,twocolumn,amsmath,amssymb,bibnotes,aps,longbibliography]{revtex4-1}

\usepackage{bm}
\usepackage[hmargin=1.5cm,vmargin=2.7cm]{geometry}
\usepackage{textcomp}
\usepackage{pifont}
\usepackage{mathptmx}
\usepackage{graphicx}
\usepackage{enumerate}
\usepackage{titlesec}
\usepackage{color}
\usepackage{isotope}
\usepackage[normalem]{ulem}
\usepackage{multirow,bigdelim}
\usepackage{pifont}
\usepackage{xcolor}\definecolor{link}{RGB}{45,48,146}
\usepackage{hyperref}
\hypersetup{
pdftitle={Quantum Teleportation between Remote Qubit Memories with Only a Single Photon as a Resource},
pdfauthor={S. Langenfeld, S. Welte, L. Hartung, S. Daiss, P. Thomas, O. Morin, E. Distante, and Gerhard Rempe},
colorlinks=true, linkcolor=[RGB]{45,48,146}, citecolor=[RGB]{45,48,146}, filecolor=[RGB]{45,48,146}, urlcolor=[RGB]{45,48,146}}

\newcommand{\ket}[1]{|{#1}\rangle}
\newcommand{\braket}[1]{\langle{#1}\rangle}
\DeclareTextFontCommand{\emph}{\textit}
\makeatletter
\@ifpackageloaded{hyperref}{\newcommand{\positionlabel}[1]{\hypertarget{#1}{}}}{\newcommand{\positionlabel}[1]{}}
\@ifpackageloaded{hyperref}{}{\newcommand{\hyperref}[2][]{#2}}
\@ifpackageloaded{hyperref}{}{}
\makeatother

\begin{document}

\title{Quantum Teleportation between Remote Qubit Memories with Only a Single Photon as a Resource}

\author{Stefan~Langenfeld$^{1}$}
\thanks{S.L.\ and S.W.\ contributed equally to this work.}
\author{Stephan~Welte$^{1}$}
\email[To whom correspondence should be addressed. Email: ]{stephan.welte@mpq.mpg.de}
\author{\hspace{-.4em}\textcolor{link}{\normalfont\textsuperscript{$*$}}\hspace{.4em}Lukas~Hartung$^{1}$}
\author{Severin~Daiss$^{1}$}
\author{Philip~Thomas$^{1}$}
\author{Olivier~Morin$^{1}$}
\author{Emanuele~Distante$^{1}$}
\author{Gerhard~Rempe$^{1}$}
\affiliation{$^{1}$Max-Planck-Institut f{\"u}r Quantenoptik, Hans-Kopfermann-Strasse 1, 85748 Garching, Germany}

\begin{abstract}
\noindent Quantum teleportation enables the deterministic exchange of qubits via lossy channels. While it is commonly believed that unconditional teleportation requires a preshared entangled qubit pair, here we demonstrate a protocol that is in principle unconditional and requires only a single photon as an ex-ante prepared resource. The photon successively interacts, first, with the receiver and then with the sender qubit memory. Its detection, followed by classical communication, heralds a successful teleportation. We teleport six mutually unbiased qubit states with average fidelity $\overline{\text{F}}=(88.3\pm1.3)\%$ at a rate of $6\,\mathrm{Hz}$ over $60\,\mathrm{m}$.\\
\end{abstract}

\maketitle 
\noindent The direct transfer of a qubit over a long distance constitutes a fundamental problem due to unavoidable transportation losses in combination with the no-cloning theorem \cite{wootters1982}. A solution was provided by Bennett et al. who brought forward the idea of quantum teleportation in a seminal paper in 1993 \cite{Bennett1993}. Here, a sender `Alice' owns a precious and unknown input qubit that she wants to communicate to a receiver `Bob'. The two are connected via a lossy quantum channel and a deterministic classical channel. In a first step, Alice and Bob use the probabilistic quantum channel to share an entangled pair of particles with a repeat-until-success strategy. Once a classical signal heralds the availability of this entanglement resource, Alice performs a joint Bell-state measurement (BSM) on the input qubit and her half of the entangled pair and communicates the outcome classically to Bob. By applying a unitary rotation conditioned on this result to his half of the resource pair, he can eventually recover the state of the input qubit. Soon after the theoretical proposal of this protocol, first proof of principle experiments were performed with photons \cite{bouwmeester1997,boschi1998, furusawa1998} and later extended to other platforms such as atoms \cite{nolleke2013}, ions \cite{riebe2004, barrett2004, olmschenk2009}, nitrogen vacancy centers \cite{pfaff2014}, atomic ensembles \cite{bao2012, krauter2013}, and hybrid systems \cite{takeda2013, bussieres2014}.

The complete teleportation protocol poses however some demanding experimental challenges \cite{pirandola2015}. First, Alice must keep the input qubit alive while the entanglement is generated over the quantum channel. This requires the qubit to be stored in a long-lived quantum memory. Second, the entanglement distribution over the lossy channel must be heralded. This allows Alice to perform the BSM only when the link is ready, avoiding the waste of her precious qubit. For the same reason, a deterministic BSM that can faithfully distinguish all four Bell states must be implemented. In earlier attempts of teleportation between distant material qubits \cite{olmschenk2009, bao2012, nolleke2013} the BSM was intrinsically probabilistic and the entanglement distribution was not heralded. Practically unconditional teleportation where Alice's input qubit always reappears on Bob's side was later reported in Ref. \cite{pfaff2014}. In this experiment, an additional ancillary matter qubit was employed to independently herald the availability of the entanglement resource before a deterministic BSM was performed, very much in the spirit of Ref. \cite{Bennett1993}.

Here we offer an alternative solution and demonstrate a novel teleportation protocol that allows for, in principle, unconditional teleportation without the necessity of the commonly employed pre-shared entanglement resource. Instead, the only resource needed prior to the start of the teleportation procedure is a single photon traveling from Bob to Alice. If the photon is lost on the way, Alice's qubit is not affected and the protocol can simply be repeated until a successful photon transmission is heralded with downstream photodetectors. Instead of presharing the entanglement resource, the entanglement is generated on the fly between Bob's qubit and the photon when the latter interacts with his node. Notably, this entanglement generation process is in principle deterministic. Furthermore, our scheme implements a BSM with no fundamental efficiency limitation. The successful detection of the photon at Alice's node excludes events of photon loss and acts as a herald both for the entanglement generation on Bob's side and for the BSM on Alice's side. Our teleportation protocol practically achieves high teleportation rates and fidelities, and is ideally suited for future quantum networks.

\begin{figure}[htb]
\centering
\includegraphics[width=\columnwidth]{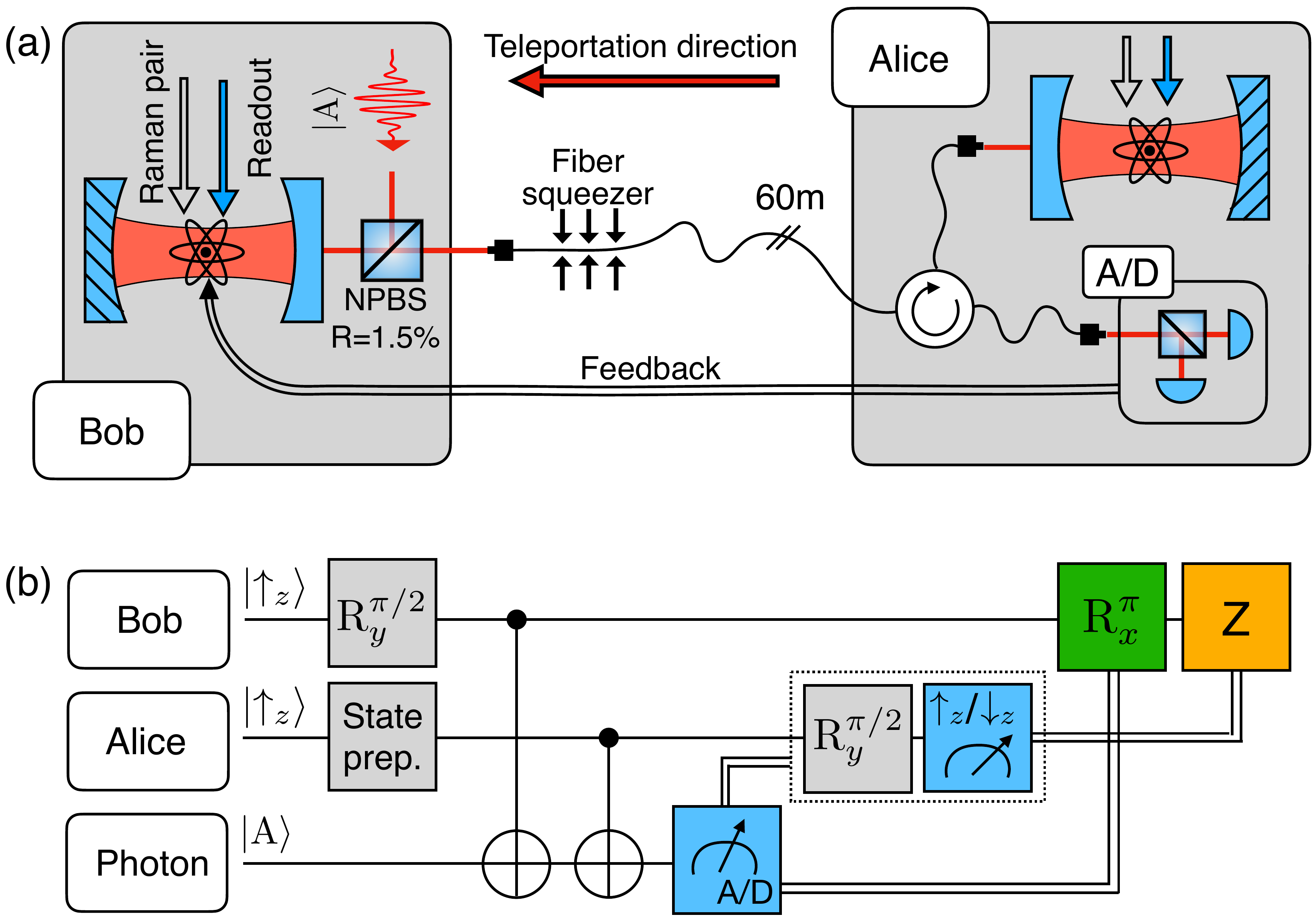}
\caption{\label{fig:setup} Teleportation of a qubit from Alice to Bob. (a) Setup: Two cavity QED setups are connected with a 60m long optical fiber. The respective atomic qubits are controlled using a pair of Raman lasers (gray arrows) and read out using a different laser beam (blue arrows). A photon (red wiggly arrow) impinges onto Bob's setup, is reflected and then propagates to Alice's setup. A fiber circulator (circular arrow) directs the photon onto Alice's cavity and, after its reflection, to a polarization-resolving detection setup. Feedback (double arrow) acts on Bob's atomic qubit. (b) Quantum circuit diagram. Single-qubit rotations are labeled with an R. The superscript describes the respective pulse area while the subscript describes the rotation axis x ($\ket{\uparrow_z}+\ket{\downarrow_z}$) or y ($\ket{\uparrow_z}+i\ket{\downarrow_z}$). After state preparation of the atoms, the photon performs two controlled-NOT gates. To make the protocol unconditional, the $\pi/2$ rotation and measurement of Alice's qubit (dashed box) could be applied only in case of a successful photon detection event independent of the polarization. At the end of the protocol, a state detection of Alice's atom and the photon is performed (blue squares). These measurement outcomes determine the two classical feedback signals.}
\end{figure}

Figure \ref{fig:setup} (a) shows a sketch of our experimental setup.
We employ two \isotope[87]{Rb} atoms trapped in two high-finesse optical cavities that are physically separated by 21m. The two atom-cavity systems are connected with a 60m long single mode optical fiber. Each of the two atoms carries one qubit of information encoded in the states $\ket{5\isotope[2]{S}_{1/2},F=2, m_F=2}:=\ket{\uparrow_z}$ and $\ket{5\isotope[2]{S}_{1/2},F=1, m_F=1}:=\ket{\downarrow_z}$. Both cavities are actively tuned to the atomic resonance $\ket{\uparrow_z}\leftrightarrow\ket{\text{e}}:=\ket{5\isotope[2]{P}_{3/2},F'=3, m_F=3}$. On this particular transition, the atom-cavity systems operate in the strong-coupling regime. 

To teleport a qubit state from Alice to Bob, a single photon is successively reflected from the two quantum nodes, starting on Bob's side.
After the second reflection, a circulator is used to direct the light to a polarization resolving setup of superconducting nanowire photon detectors. Given the photon click, an atomic state detection is performed on Alice's qubit employing a laser beam resonant with the $\ket{\uparrow_z}\leftrightarrow\ket{e}$ transition. The respective state-detection light is directed into the same detectors as the reflected photon. Depending on the outcome of both the atomic and the photonic measurements, feedback pulses are applied to Bob's qubit. To benchmark this teleportation protocol, we perform a full state tomography of Bob's qubit by measuring the atomic state in three mutually orthogonal bases. The respective scattered state-detection light is directed to an additional tomography setup 
of single-photon detectors (not shown in Fig. \ref{fig:setup}(a)).
\label{sec:protocol}\\

We remark that the fundamental building block of our protocol is an atom-photon controlled-NOT gate that is executed by reflecting a photon from the atom-cavity system \cite{duan2004, reiserer2014}. It is based on a polarization flip of the photon from antidiagonal (diagonal) polarization to the orthogonal diagonal polarization whenever the atom is in the state $\ket{\uparrow_z}$. In the case of a noncoupling atom in $\ket{\downarrow_z}$, the polarization of the photon does not change. The full teleportation protocol then consists of a photon performing a controlled NOT gate at each node in succession, followed by different conditional feedback pulses. 

The quantum circuit diagram of our full scheme is shown in Fig. \ref{fig:setup}(b). We start by optically pumping both atoms to the initial state $\ket{\uparrow_z}$. Then, Alice's atom is initialized to a general state of the form $\alpha\ket{\uparrow_z}+\beta\ket{\downarrow_z}$. To this end, we employ a two-photon Raman process that allows us to initialize any desired qubit state by an appropriate choice of amplitude, phase, and timing. Bob's atom is prepared in the equal-superposition state $(\ket{\uparrow_z}+\ket{\downarrow_z})/\sqrt{2}$.
We use a very weak coherent pulse (average photon number $\braket{n}=0.07(1)$) in combination with single-photon detectors as an approximation of a heralded single-photon source. The light is initially prepared in the antidiagonal polarization $\ket{A}$.
This pulse is first reflected from Bob's cavity, creating a maximally entangled state between the photon and his atom.
By generating the entanglement on the fly during the reflection, far-away Alice is not yet involved in the teleportation protocol. Subsequently, we reflect the light from Alice's cavity. The resulting atom-atom-photon state can be expressed as (see Supplemental Material \cite{supplement})
\begin{equation}\frac{1}{\sqrt{2}}\Big[\Big(\alpha\ket{\uparrow_z\uparrow_z}+\beta\ket{\downarrow_z\downarrow_z}\Big)\ket{\text{A}}+\Big(\beta{\ket{\uparrow_z\downarrow_z}}+\alpha\ket{\downarrow_z\uparrow_z}\Big)\ket{\text{D}}\Big].\label{eq:threetangle}\end{equation}
Notably, the light polarization has changed from $\ket{\text{A}}$ to $\ket{\text{D}}$ whenever an odd number of coupling atoms is present in the two nodes, resulting in a three-particle entangled state. The presence of a photon in the weak coherent pulse is then heralded with single-photon detectors that also register the polarization of the light and thus project the combined state in Eq. (\ref{eq:threetangle}). Now a $\pi/2$ pulse is applied on Alice's atom and then its state is measured. The outcome of the atomic state detection is analyzed and a phase gate is executed on Bob's qubit whenever the result of the state detection is $\ket{\uparrow_z}$. A further feedback is applied depending on the light's measured polarization state. For a detection in $\ket{\text{D}}$, a $\pi$ pulse around the $x-$axis is executed. It inverts the role of $\ket{\uparrow_z}$ and $\ket{\downarrow_z}$ without introducing a relative phase between them. This second feedback completes the teleportation protocol and the state of Bob's atom is $\alpha\ket{\uparrow_z}+\beta\ket{\downarrow_z}$.
Excluding optical pumping ($200\,\mathrm{\mu s}$), the entire protocol takes $25.5\,\mathrm{\mu s}$. This is currently limited by the duration of the Raman pulses ($4\,\mathrm{\mu s}$ for a $\pi/2$ pulse).
Afterwards we apply a cooling sequence to the atom. We set the repetition rate of the experiment to $1\,\mathrm{kHz}$. The probability for a successful transmission of a single photon through the entire setup and an eventual detection amounts to $8.4\%$ \cite{supplement}. Therefore, employing a single-photon source would yield a teleportation rate of $84\,\mathrm{Hz}$. However, due to the additional use of weak coherent pulses ($\braket{n}=0.07$), it is reduced to $6\,\mathrm{Hz}$ in our implementation.\\ 

\begin{figure*}[t] 
\includegraphics[width=1.8\columnwidth]{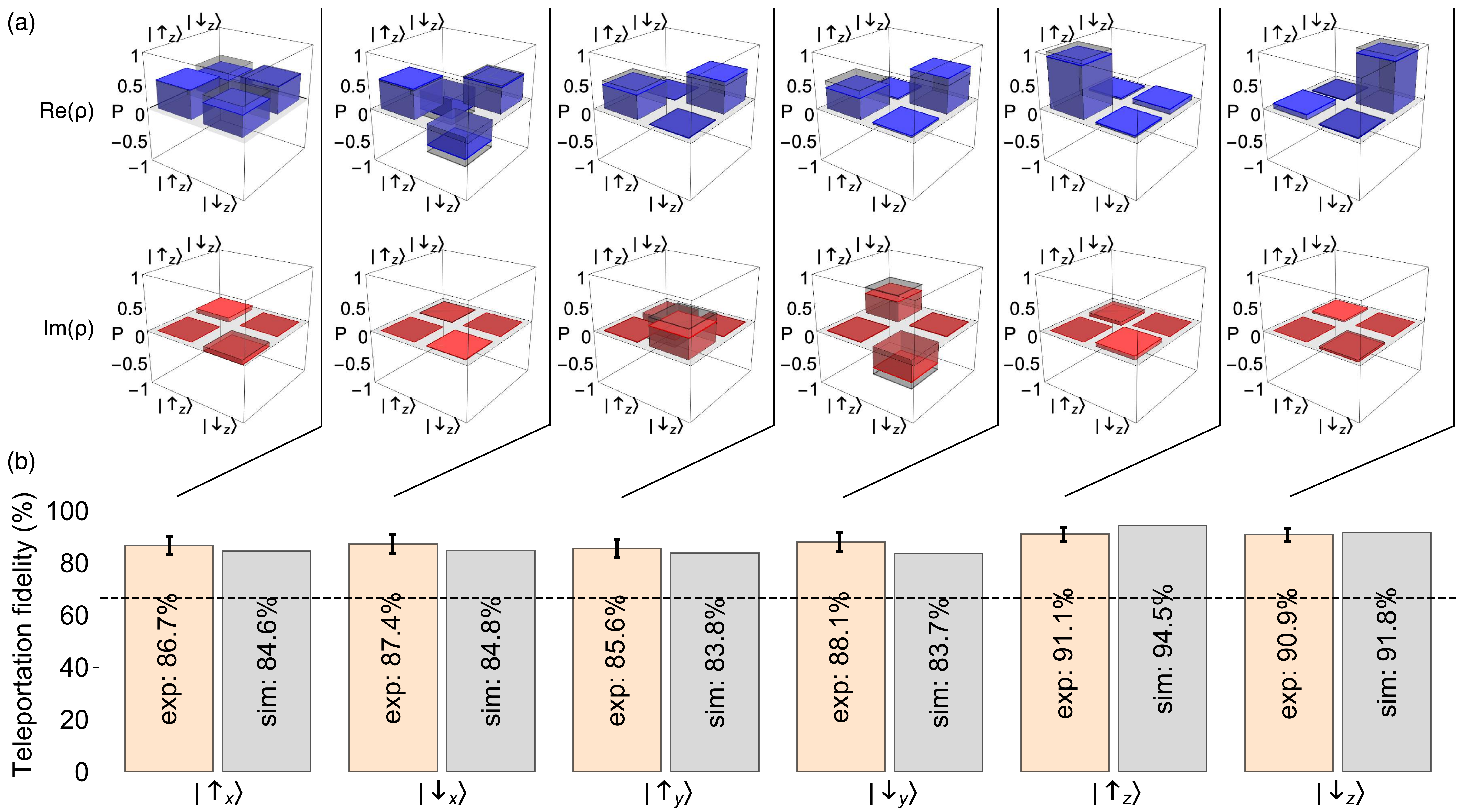}
\caption{\label{fig:results} Teleportation results. (a) Density matrices of the single-qubit states that were teleported from Alice to Bob. (b) Teleportation fidelities of the six teleported states. The orange bars show the experimentally (exp) measured fidelities. The error bars are standard deviations of the mean. In gray, we show the simulated (sim) fidelities. The dashed line represents the classically achievable threshold of $2/3$ which can be reached with a measure-prepare strategy \cite{massar1995}.}
\end{figure*}    

To characterize the performance of our teleportation protocol, we prepare Alice's qubit in six mutually unbiased states and teleport these states to Bob. The six states are $\ket{\uparrow_z}$, $\ket{\downarrow_z}$, $\ket{\uparrow_x}:=1/\sqrt{2}(\ket{\uparrow_z}+\ket{\downarrow_z})$, $\ket{\downarrow_x}:=1/\sqrt{2}(\ket{\uparrow_z}-\ket{\downarrow_z})$, $\ket{\uparrow_y}:=1/\sqrt{2}(\ket{\uparrow_z}+i\ket{\downarrow_z})$, and $\ket{\downarrow_y}:=1/\sqrt{2}(\ket{\uparrow_z}-i\ket{\downarrow_z})$. After the teleportation, a quantum state tomography of Bob's atom is performed to extract its complete density matrix. The results for the six prepared input states are shown in Fig. \ref{fig:results}(a). All teleported states have a high overlap with the ideally expected result. On average, we achieve a teleportation fidelity of $(88.3\pm1.3)\%$ which is significantly higher than the classically achievable threshold of $2/3$ \cite{massar1995}. 

To understand the limitations in the achieved fidelities, we simulate our experiment with the quantum optics toolbox QuTiP \cite{johansson2012}. The simulated fidelities are depicted as the gray bars in Fig. \ref{fig:results}(b) and show an excellent agreement with the experiment. From this we conclude that the fidelity of the teleported states is mainly influenced by three sources of error, namely, qubit decoherence (only when the atoms are in superposition states), two-photon contributions in the coherent laser pulses, and imperfect atomic state preparation. These three effects reduce the fidelity by $6.0\%$, $3.9\%$, and $1.4\%$, respectively. Additional imperfections like mechanical vibrations of the cavity mirrors, imperfect fiber birefringence compensation, and polarization dependent losses are minor errors that sum up to the rest of the accumulated infidelity. For a description of the simulation model, see the Supplemental Material \cite{supplement}.

By protocol, our teleportation scheme is designed for employing a single photon to be reflected from the two nodes. Nevertheless, the teleportation can also be executed with weak coherent pulses which, in practice, are much easier to produce than single photons. Although the use of a coherent pulse is not ideal as residual higher photon number contributions in the coherent pulse deteriorate the atom-photon entanglement \cite{reiserer2014} and thus the teleportation fidelity, the teleportation rate can be increased simply by increasing $\braket{n}$. Conversely, in the limit of vanishing $\braket{n}$, the teleportation fidelity approaches the scenario where a single-photon source is employed. To characterize this, we performed an additional measurement where the dependence of the teleportation fidelity on $\braket{n}$ was investigated. For this measurement, Alice's atom is prepared in the state $\ket{\uparrow_x}$, which is most sensitive to experimental imperfections. Afterwards, the teleportation protocol is executed. Figure \ref{fig:alpha_scan} shows the obtained data and an expected curve based on the simulation outlined in the Supplemental Material \cite{supplement}.
\begin{figure}[t]
\centering
\includegraphics[width=\columnwidth]{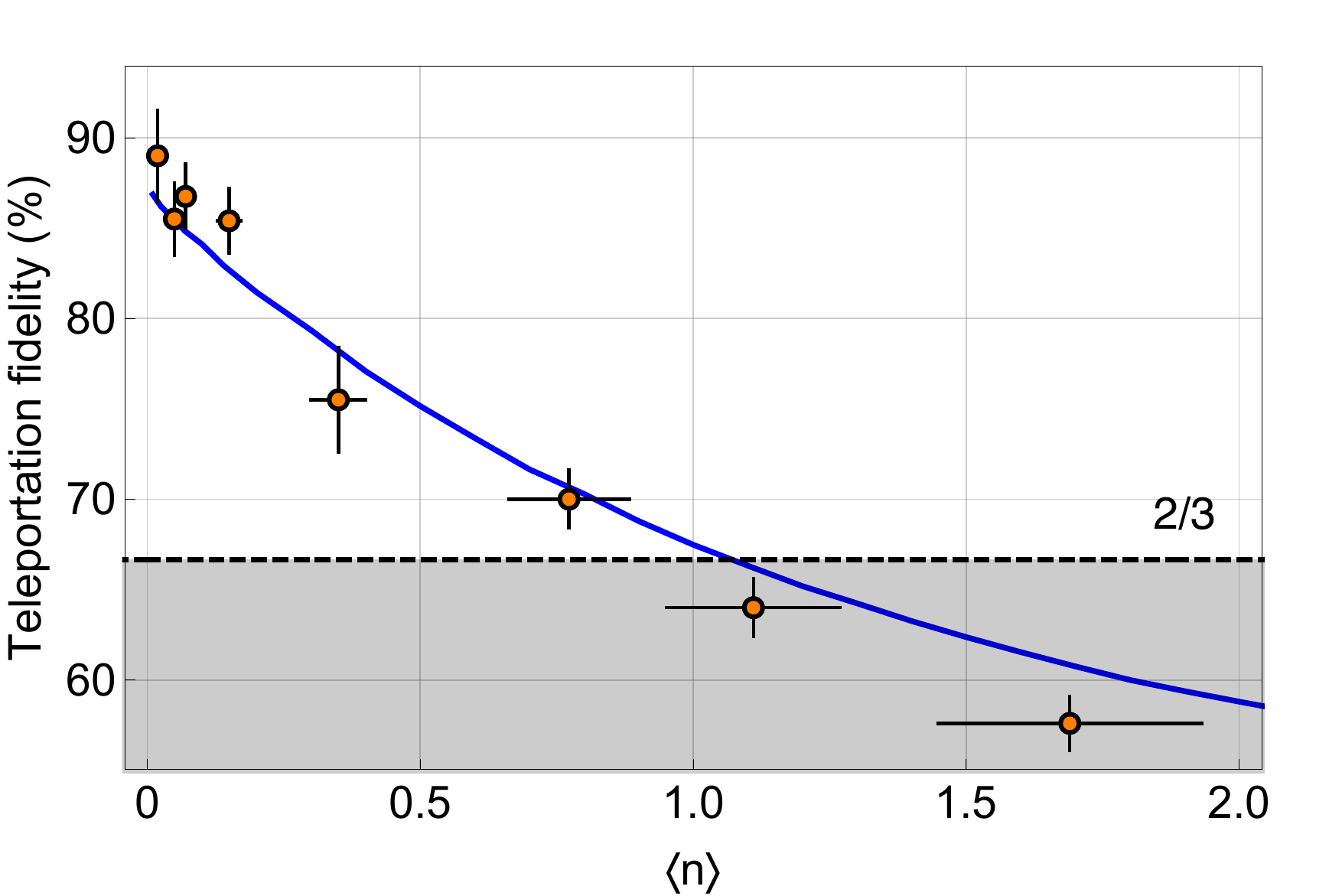}
\caption{\label{fig:alpha_scan} Scan of mean photon number. Teleportation fidelity as a function of the mean photon number in the employed coherent pulse. The blue line is based on a theoretical model (see Supplemental Material \cite{supplement}). The dashed line shows the classical teleportation threshold of 2/3. The error bars represent standard deviations from the mean.}
\end{figure}
As expected, the teleportation fidelity yields its highest values for a vanishing mean photon number $\braket{n}$ and decreases due to the higher photon-number contributions when it is increased. For the smallest employed value of $\braket{n}=0.02$, the fidelity reaches $89\%$. Our measurements show that the classical threshold of $2/3$ is beaten up to $\braket{n}\approx1.0$. At this photon number we achieve a teleportation rate of $84\,\mathrm{Hz}$, more than 1 order of magnitude higher than in the case of our default employed photon number of $\braket{n}=0.07$ (see Supplemental Material \cite{supplement} for more information about the teleportation rate).

\begin{figure}[t]
\includegraphics[width=1.0\columnwidth]{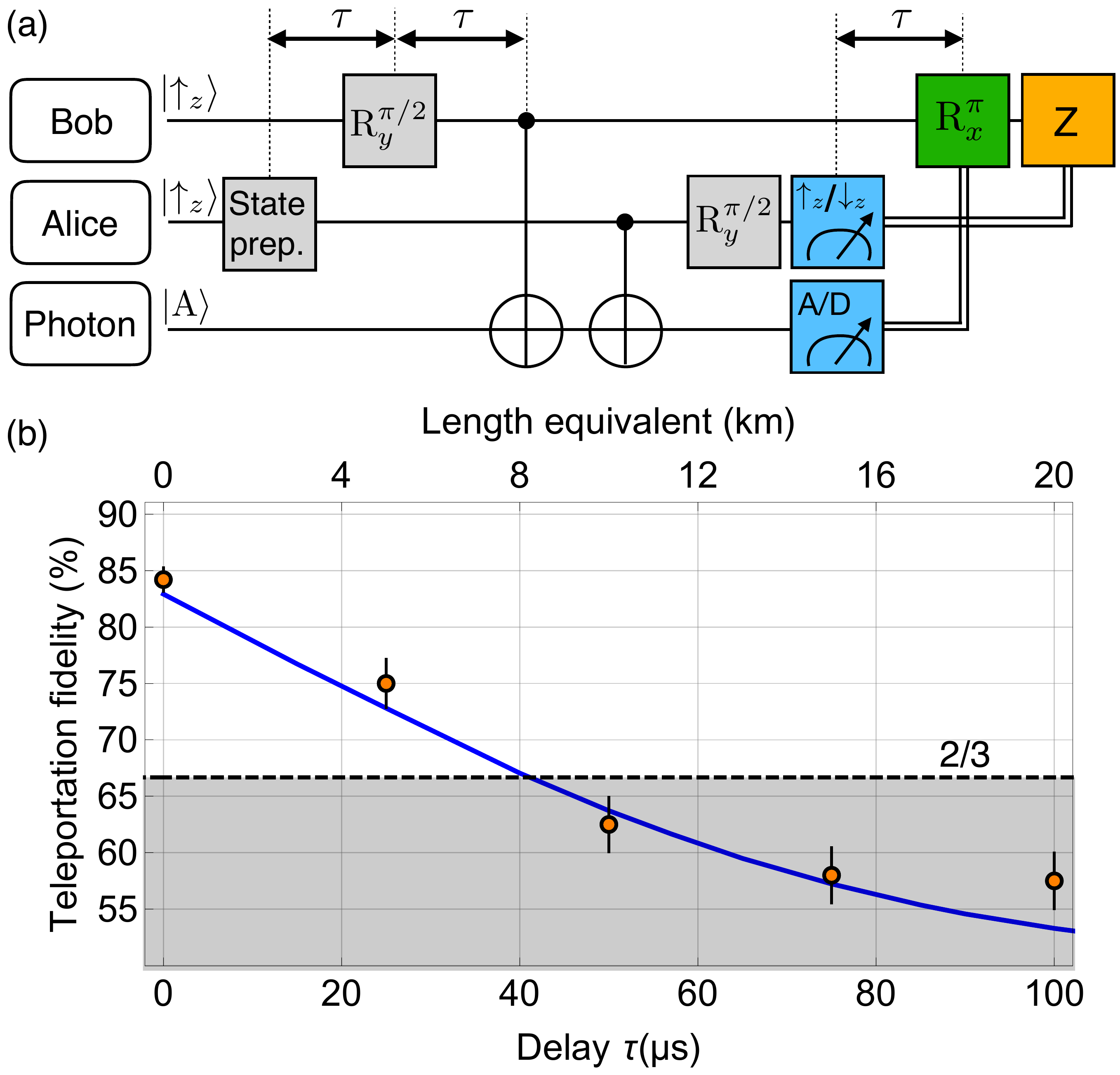}
\caption{\label{fig:delay_scan} Scan of the delay. (a) Quantum circuit diagram with the added three delay intervals $\tau$. (b) Teleportation fidelity versus the variable delay. The dashed line represents the classical threshold of 2/3. The blue line is based on our theoretical model (see Supplemental Material \cite{supplement}). The upper horizontal axis represents the length equivalent corresponding to $\tau\times \mathrm{c}/1.5$, where c is the speed of light and 1.5 the refractive index of the fiber. In this experiment, the mean photon number was chosen a factor of 2 higher compared to the data shown in Fig. \ref{fig:results}.}
\end{figure}

To mimic a variable distance between the two network nodes, we introduce a variable temporal delay $\tau$ at different times in our teleportation protocol. The modified version of the protocol is shown in Fig. \ref{fig:delay_scan}(a).
First, we introduce a delay $\tau$ between the two state-preparation pulses to simulate a communication time after Alice's qubit is prepared. Afterwards, an additional $\tau$ mimics a longer propagation time of the photon in the fiber.
Eventually a third delay takes into account the propagation time of the two feedback signals. In Fig. \ref{fig:delay_scan}(b), the teleportation fidelity of Alice's state $\ket{\uparrow_x}$ is plotted against the respective delay $\tau$. The qubits in the two nodes are both in superposition states.
Thus, the increase of $\tau$ increases the effect of atomic decoherence and thereby reduces the achieved teleportation fidelities. Our measurements show that the classical threshold fidelity is beaten up to a delay of $\tau\approx40\,\mathrm{\mu s}$. This delay corresponds to a fiber length of $8\,\mathrm{km}$, a range comparable to an urban quantum network. It should be noted, however, that our measurements neither simulate the additional fiber fluctuations in an urban environment nor the additional losses in a fiber that is physically longer. 

To conclude, we have devised and implemented a novel and, in principle, unconditional scheme to perform quantum teleportation in a network. An advantage of our protocol is that the underlying photon-reflection mechanism is robust with respect to the temporal shape of the employed photon as was demonstrated in Ref. \cite{daiss2019}. This makes our scheme easy to implement in comparison to the conventionally employed overlapping of two identical photons on a beam splitter \cite{olmschenk2009, bao2012, nolleke2013, meraner2020}. In combination with the scheme demonstrated in Ref. \cite{kalb2015}, and in contrast to Ref. \cite{pfaff2014}, our protocol would allow for teleportation between any combination of unknown matter and light qubits. It would even be possible to convert \cite{huang1992} the wavelength of the ancilla photon during its passage from Bob to Alice in case the two communication partners employ different kinds of qubits. Furthermore, in an improved setup with cavities having a reflectivity close to unity and negligible photon loss between Alice's cavity and the downstream detectors \cite{bhaskar2020}, our implementation of teleportation would be unconditional. In the current realization, this is not yet the case since the loss of a photon could happen after the interaction with Alice's qubit and therefore damage it. Lastly, our protocol is platform independent and could be implemented with different carriers of quantum information coupled to resonators such as vacancy centers in diamond \cite{bhaskar2020}, rare-earth ions \cite{chen2020}, superconducting qubits \cite{besse2018, kono2018}, or quantum dots \cite{fushman2008, desantis2017, sun2018}. 

\begin{acknowledgments}
This work was supported by the Bundesministerium f\"{u}r Bildung und Forschung via the Verbund Q.Link.X (16KIS0870), by the Deutsche Forschungsgemeinschaft under Germany’s Excellence Strategy – EXC-2111 – 390814868, and by the European Union’s Horizon 2020 research and innovation programme via the project Quantum Internet Alliance (QIA, GA No. 820445). E.D. acknowledges early support by the Cellex-ICFO-MPQ postdoctoral fellowship program.
\end{acknowledgments}

\clearpage

\section{SUPPLEMENTAL MATERIAL}
\label{supplement}
\section{Simulation Model Implemented with QuTip}
We use the python toolkit QuTip \cite{johansson2012supplement} to simulate our teleportation experiment. The underlying theoretical modelling is based on a cavity input-output theory adapted from \cite{kuhn2015, hacker2019}. In total, we consider four possible scattering modes of a photon after the interaction with the network node. These modes are the cavity reflection, the cavity transmission, scattering on the mirrors and the scattering via the atom. The physical system is modelled with a concatenation of four beam splitters each corresponding to one of the four possible output modes of the system. The reflection of the cavity is additionally equipped with a phase shift operator to take into account the phase shift occurring in the reflection process \cite{duan2004supplement}. The evolution of the initial atom-light state is governed by the operator built from the four beam splitters and the phaseshift operation. In the end the atom scattering, the mirror losses and the cavity transmission are traced out to obtain the result for the desired reflection mode of the cavity.\\
In the experiment, we have two cavity systems connected with an optical fiber that acts as a quantum channel. This channel is modelled in the theory as a retarder and a beamsplitter that describe a lossy depolarizing channel. \\
Eventually, the photons are measured with single-photon detectors. We simulate their non-unity efficiency ($\eta\approx90\%$ at $\lambda=780\,\mathrm{nm}$) and non-zero dark counts ($9\,\mathrm{Hz}$) with a beam splitter where one port is supplied with thermal noise. \\
The simulation evolves the atom-atom-photon state through the beam splitters describing the two reflections, the lossy depolarizing fiber and the non-perfect photon detectors. Afterwards, the projective measurement of the light polarization is included. The simulation then outputs the resulting final density matrix. The theory uses experimental parameters such as the mode matching of the photons to the cavities, the state preparation fidelities, the atomic coherence times and the dark counts of the detectors. These parameters were determined in independent characterization measurements.  
\label{subsec:supplement}
\section{Summary of the Experimental Parameters}
Here, we give a summary of the most important experimental parameters. The protocol starts by optical pumping the two atoms to the state $\ket{\uparrow_z}$. This is achieved with a fidelity of $99\%$ in either of the setups. $\pi$ pulses are performed within $8\,\mathrm{\mu s}$ and yield residual populations of $3\%$ in $\ket{\uparrow_z}$. The atomic coherence time in each of the two systems is approximately $400\,\mathrm{\mu s}$.\\ 
The cavities are actively tuned to the $\ket{\uparrow_z}\leftrightarrow\ket{e}$ transition, where the cavity QED parameters assume the values summarized in Table \ref{tab:parametersPQ}. Both systems operate in the strong coupling regime.
\begin{table}[htbp]
\centering
\begin{tabular}{|c|c|c|}\hline
parameter & node 1 (Bob) & node 2 (Alice)  \\ \hline\hline
$\kappa (\text{MHz})$ & 2.5 & 2.8  \\ 
$\gamma (\text{MHz})$ & 3.0 & 3.0 \\ 
$g (\text{MHz})$ & 7.6 & 7.6 \\ \hline 
$C$ & 3.9 & 3.4\\
\hline 
\end{tabular} 
\caption[Tabelle]{Cavity QED parameters of the two network nodes. The given parameters are the cavity linewidth $\kappa$, the atomic polarization decay-rate $\gamma$ and the atom-cavity coupling strength $g$. We define the cooperativity of the atom-cavity system as $C=g^2/(2\kappa\gamma)$.}
\label{tab:parametersPQ} 
\end{table}
Between the two setups, the light traverses various optical elements which introduce losses. The 60m fiber with the fiber circulator, color filters, mirror and lens surfaces lead to a total transmission of $51\%$ between the two resonators. Additionally, the reflectivities of the two cavity systems are $60\%$ and $55\%$ on resonance, respectively. The detection efficiency for light after the second cavity system amounts to $50\%$ and includes a second passage through the fiber circulator, incoupling and absorption losses of the respective fiber and the detection efficiency of our single-photon detectors. The total probability to detect a single photon after propagation of a weak coherent pulse (average photon number $\braket{n}=0.07$) through the entire system amounts to $6\times 10^{-3}$. This is also the overall efficiency of our teleportation protocol for this mean photon number. We run the experiment at a repetition rate of $1\,\mathrm{kHz}$, and thus achieve a teleportation rate of $6\,\mathrm{Hz}$. 

\section{Detailed Description of the Teleportation Protocol}
Here we describe the experimental protocol in more detail and calculate the individual states created at the different steps of the teleportation protocol.\\
Given that there is an atom available at each node, we first initialize them in $\ket{\uparrow_z}$. The optical pumping of Bob's atom takes $200\,\mathrm{\mu s}$ while it takes $240\,\mathrm{\mu s}$ for Alice's atom. In both network nodes, we employ a pair of Raman lasers to initialize the desired superposition states. We follow the convention of \cite{nielsen2000} and define a $\theta=\pi/2$ rotation around the $y-$axis with the rotation matrix \begin{equation}R_y(\theta)=\begin{pmatrix}\cos\theta/2 & -\sin\theta/2 \\\sin\theta/2 & \cos\theta/2\\\end{pmatrix}\overset{\theta=\pi/2}{=}\frac{1}{\sqrt{2}}\begin{pmatrix}1 & -1 \\1 & 1\\\end{pmatrix}.\end{equation} On Bob's side, this transformation is used to generate the state $\frac{1}{\sqrt{2}}(\ket{\uparrow_z}+\ket{\downarrow_z})$ while Alice prepares a general state of the form $\alpha\ket{\uparrow_z}+\beta\ket{\downarrow_z}$. Therefore the starting state of the teleportation protocol can be written as a separable state of the form
\begin{equation}
\begin{split}
&\frac{1}{\sqrt{2}}(\ket{\uparrow_z}+\ket{\downarrow_z})\otimes(\alpha\ket{\uparrow_z}+\beta\ket{\downarrow_z})\\
&=\frac{1}{\sqrt{2}}(\alpha\ket{\uparrow_z\uparrow_z}+\beta\ket{\uparrow_z\downarrow_z}+\alpha\ket{\downarrow_z\uparrow_z}+\beta\ket{\downarrow_z\downarrow_z}).
\end{split}
\end{equation}
The photon is reflected from the two cavity QED systems successively. It has a Gaussian envelope with a full width at half maximum of $1\,\mathrm{\mu s}$ and an antidiagonal polarization $\ket{\text{A}}$. Due to the atom-photon gate mechanism, a reflection yields
\begin{equation}
\begin{split}
\ket{\downarrow_z}\ket{\text{A}}&\rightarrow\ket{\downarrow_z}\ket{\text{A}}\\
\ket{\uparrow_z}\ket{\text{A}}&\rightarrow\ket{\uparrow_z}\ket{\text{D}}.
\end{split}
\end{equation}
Applying this to the reflection from Bob's cavity, the resulting atom-photon state is $\frac{1}{\sqrt{2}}(\ket{\uparrow_z}\ket{\text{D}}+\ket{\downarrow_z}\ket{\text{A}})$, a maximally entangled Bell state. This state is the required resource for our teleportation protocol. After the reflection from Alice's cavity, the combined atom-atom-photon state is then
\begin{equation}
\frac{1}{\sqrt{2}}\Big(\alpha\ket{\uparrow_z\uparrow_z}\ket{\text{A}}+\beta{\ket{\uparrow_z\downarrow_z}}\ket{\text{D}}+\alpha\ket{\downarrow_z\uparrow_z}\ket{\text{D}}+\beta\ket{\downarrow_z\downarrow_z}\ket{\text{A}}\Big).
\end{equation}
Now a $\pi/2$ rotation around the $y-$axis is applied to Alice's atom. This results in the state
\begin{align}
\frac{1}{2}
&\Big(\alpha\ket{\uparrow_z\uparrow_z}\ket{\text{A}}+\alpha\ket{\uparrow_z\downarrow_z}\ket{\text{A}}+\alpha\ket{\downarrow_z\uparrow_z}\ket{\text{D}}+\alpha\ket{\downarrow_z\downarrow_z}\ket{\text{D}} \nonumber\\
&-\beta\ket{\uparrow_z\uparrow_z}\ket{\text{D}}+\beta\ket{\uparrow_z\downarrow_z}\ket{\text{D}}-\beta\ket{\downarrow_z\uparrow_z}\ket{\text{A}}+\beta\ket{\downarrow_z\downarrow_z}\ket{\text{A}}\Big).  
\end{align}
The measurement of the polarization state of the photon in the A/D basis projects the two atom state to
\begin{align}
\frac{1}{\sqrt{2}}\Big(\alpha\ket{\uparrow_z\uparrow_z}+\alpha\ket{\uparrow_z\downarrow_z}-\beta\ket{\downarrow_z\uparrow_z}+\beta\ket{\downarrow_z\downarrow_z}\Big)
\end{align}
for the $\ket{\text{A}}$ polarization and to
\begin{align}
\frac{1}{\sqrt{2}}\Big(\alpha\ket{\downarrow_z\uparrow_z}+\alpha\ket{\downarrow_z\downarrow_z}-\beta\ket{\uparrow_z\uparrow_z}+\beta\ket{\uparrow_z\downarrow_z}\Big)
\end{align}
for the $\ket{\text{D}}$ polarization. As a last step of the protocol, the two feedback pulses are applied. For a photon detection in $\ket{\text{D}}$, a $\pi$ rotation around the x-axis is applied to Bob's atom. In matrix form it reads 
\begin{equation}
R_x(\theta)=\begin{pmatrix}\cos\theta/2 & -i\sin\theta/2 \\-i\sin\theta/2 & \cos\theta/2\\\end{pmatrix}\overset{\theta=\pi}{=}\begin{pmatrix}0 & -i \\-i & 0\\\end{pmatrix}.
\end{equation}
Up to a global phase, this transformation inverts the roles of $\ket{\uparrow_z}$ and $\ket{\downarrow_z}$. As a result of this feedback, the two-atom state becomes
\begin{equation}
\frac{1}{\sqrt{2}}\Big(\alpha\ket{\uparrow_z\uparrow_z}+\alpha\ket{\uparrow_z\downarrow_z}-\beta\ket{\downarrow_z\uparrow_z}+\beta\ket{\downarrow_z\downarrow_z}\Big).
\end{equation}
For the second feedback, we measure the state of Alice's atom. If the result is $\ket{\uparrow_z}$, a Z-gate (phase gate) is applied to Bob's atom. For $\ket{\downarrow_z}$, no feedback is applied. In both cases, the resulting state on Bob's side in either case is $\alpha\ket{\uparrow_z}+\beta\ket{\downarrow_z}$ and the teleportation is complete. Table \ref{tab:feedback} shows an overview of the different measurement outcomes and the different feedback signals. \\ \\
\begin{table}[htb]
\centering
 \begin{tabular}{|c|c|} 
 \hline
Measured state photon / Alice's atom & Feedback on Bob's atom  \\ [0.5ex] \hline\hline
$\ket{\text{A}}$& none\\ 
 \hline
 $\ket{\text{D}}$ &$R_x(\pi)$ \\
 \hline
 $\ket{\uparrow_z}$& Z-gate\\
 \hline
 $\ket{\downarrow_z}$& none\\ 
\hline
\end{tabular}
\caption[Tabelle]{Feedback used in the teleportation protocol. The table lists the feedback pulses that are applied to Bob's atom for certain measured states of the photon and Alice's atom.} 
\label{tab:feedback}
\end{table}

 \clearpage
\end{document}